\documentclass[sigconf]{acmart}

\usepackage{algorithm}
\usepackage{algorithmic}
\usepackage{amsmath}
\usepackage{booktabs}
\usepackage{adjustbox}

\usepackage{array}
\usepackage{hhline}
\usepackage{graphicx}
\usepackage{subcaption}
\usepackage{stfloats}
\usepackage{float}
\usepackage{enumitem}
\usepackage{xcolor}

\usepackage{caption}
\AtBeginDocument{%
  }

\setcopyright{acmlicensed}
\copyrightyear{2018}
\acmYear{2018}
\acmDOI{XXXXXXX.XXXXXXX}
\acmConference[Conference acronym 'XX]{Make sure to enter the correct
  conference title from your rights confirmation email}{June 03--05,
  2018}{Woodstock, NY}
\acmISBN{978-1-4503-XXXX-X/2018/06}

\begin{document}

\title{Killing Two Birds with One Stone: Unifying Retrieval and Ranking with a Single Generative Recommendation Model}

\author{Luankang Zhang}
\email{zhanglk5@mail.ustc.edu.cn}
\orcid{0009-0006-5833-5999}
\affiliation{
  \institution{University of Science and Technology of China}
  \city{Hefei}
  \state{Anhui}
  \country{China}
}

\author{Kenan Song}
\email{songkenan@huawei.com}
\orcid{0009-0004-3375-6928}
\affiliation{
  \institution{Huawei Noah’s Ark Lab}
  \city{Singapore}
  \state{Singapore}
  \country{Singapore}
}

\author{Yi Quan Lee}
\email{leeyiquan95@gmail.com}
\orcid{0009-0007-9747-741X}
\author{Wei Guo}
\email{guowei67@huawei.com}
\orcid{0000-0001-8616-0221}
\affiliation{
  \institution{Huawei Noah’s Ark Lab}
  \city{Singapore}
  \state{Singapore}
  \country{Singapore}
}


\author{Hao Wang}
\authornote{Corresponding author.}
\email{wanghao3@ustc.edu.cn}
\orcid{0000-0001-9921-2078}
\affiliation{
  \institution{University of Science and Technology of China}
  \city{Hefei}
  \state{Anhui}
  \country{China}
}

\author{Yawen Li}
\email{warmly0716@126.com}
\orcid{0000-0003-2662-3444}
\affiliation{
  \institution{Beijing University of Posts and Telecommunications}
  \city{Beijing}
  \state{Beijing}
  \country{China}
}

\author{Huifeng Guo}
\email{huifeng.guo@huawei.com}
\orcid{0000-0002-7393-8994}
\affiliation{
  \institution{Huawei Noah’s Ark Lab}
  \city{Shenzhen}
  \state{Shenzhen}
  \country{China}
}

\author{Yong Liu}
\email{liu.yong6@huawei.com}
\orcid{0000-0001-9031-9696}
\affiliation{
  \institution{Huawei Noah’s Ark Lab}
  \city{Singapore}
  \state{Singapore}
  \country{Singapore}
}

\author{Defu Lian}
\email{liandefu@ustc.edu.cn}
\orcid{0000-0002-3507-9607}
\affiliation{
  \institution{University of Science and Technology of China}
  \city{Hefei}
  \state{Anhui}
  \country{China}
}

\author{Enhong Chen}
\email{cheneh@ustc.edu.cn}
\orcid{0000-0002-4835-4102}
\affiliation{
  \institution{University of Science and Technology of China}
  \city{Hefei}
  \state{Anhui}
  \country{China}
}

\renewcommand{\shortauthors}{Luankang Zhang et al.}

\begin{abstract}
    In recommendation systems, the traditional multi-stage paradigm, which includes retrieval and ranking, often suffers from information loss between stages and diminishes performance. Recent advances in generative models, inspired by natural language processing, suggest the potential for unifying these stages to mitigate such loss. This paper presents the \textbf{Uni}fied \textbf{G}enerative \textbf{R}ecommendation \textbf{F}ramework (\textbf{UniGRF}), a novel approach that integrates retrieval and ranking into a single generative model. By treating both stages as sequence generation tasks, UniGRF enables sufficient information sharing without additional computational costs, while remaining model-agnostic.
    To enhance inter-stage collaboration, UniGRF introduces a ranking-driven enhancer module that leverages the precision of the ranking stage to refine retrieval processes, creating an enhancement loop. Besides, a gradient-guided adaptive weighter is incorporated to dynamically balance the optimization of retrieval and ranking, ensuring synchronized performance improvements.
    Extensive experiments demonstrate that UniGRF significantly outperforms existing models on benchmark datasets, confirming its effectiveness in facilitating information transfer. Ablation studies and further experiments reveal that UniGRF not only promotes efficient collaboration between stages but also achieves synchronized optimization. UniGRF provides an effective, scalable, and compatible framework for generative recommendation systems.
\end{abstract}




\maketitle
\section{Introduction}
Recommendation systems aim to identify users' potential preferences by analyzing their historical interaction behavior and selecting items they may like from the item pool~\cite{chen2020sequence,cho2014learning,rendle2010factorizing}. In practical industrial applications, due to efficiency constraints, the recommendation system is typically divided into two stages: first, a low-precision retrieval model efficiently selects a small number of items from the entire pool as a candidate set~\cite{huang2013learning,zhang2023rethinking}; then, a ranking model makes accurate predictions of user preferences within these candidate sets. Both the retrieval and ranking stages are of significant interest to academia and industry~\cite{wang2020cold,ma2021tradeoff}.

In recent years, inspired by natural language processing, the autoregressive paradigm has been adopted in recommendation systems, using users' historical behavior to predict their next actions. GRU4Rec~\cite{gru4rec} employs recurrent neural networks (RNNs)~\cite{rnn} with gated recurrent units (GRUs)~\cite{gru} to capture short-term user preferences.
SASRec~\cite{sasrec} introduces a Transformer-based~\cite{attention} model using self-attention to capture dynamic interest patterns. Although these methods improve performance, scalability remains a challenge, as larger models do not always yield better results.
To address this, newer models explore scaling laws and generative recommendation systems. 
LSRM\cite{lsrm} adds layer-by-layer adaptive dropout and switching optimizers to the SASRec framework, facilitating model scaling.
Meta's HSTU~\cite{zhai2024actions} introduces pointwise aggregated attention and the M-FALCON algorithm, expanding model dimensions to GPT-3~\cite{floridi2020gpt} levels. MBGen~\cite{MBGen} reformulates tasks to generate items based on behavior type and uses a position-routed sparse architecture for efficient scaling.
HLLM~\cite{chen2024hllm} proposes a two-layer framework where Item LLM integrates text features and User LLM predicts future interests from historical interactions. 
These generative models demonstrate that autoregressive paradigms can effectively leverage scaling laws to significantly enhance recommendation performance.

Despite significant advancements in generative recommendation models, current approaches are often limited to single-stage applications or require separate training for each stage, leading to information loss during inter-stage transmission. This loss is inherent due to training distinct generative models for each stage. While some frameworks using traditional models have attempted multi-stage joint modeling, they mainly focus on transferring information across stages from a data~\cite{cascadedata1,cascadedata2,cascadedata3} or loss~\cite{cascadeloss1,cascadeloss2} perspective, which does not effectively address data bias and information loss.
Inspired by the success of large language models (LLMs) in handling multiple tasks simultaneously~\cite{touvron2023llama,radford2019language,brown2020language}, we propose a unified generative recommendation framework that integrates multi-stage processes. This approach allows a large generative model to handle multiple stages concurrently, enabling mutual reinforcement among stages. By adopting this unified paradigm, we can inherently address information loss and facilitate sufficient information sharing across different stages.

While generative recommendation systems have the potential to unify retrieval and ranking stages into a single model, significant challenges remain. Research on unified generative frameworks is still lacking. We identify two key issues: (1). The retrieval stage filters a small subset of potentially relevant items from a large pool, anticipating that the ranking stage will assign higher scores to these selected items. Thus, retrieval and ranking are inherently interdependent. Therefore, \textbf{designing an efficient collaboration mechanism that leverages these dependencies to enhance both stages within the same generative model is a critical challenge.} (2). Empirically, different stages use different loss functions, leading to discrepancies in loss magnitude and convergence rates. As shown in Fig.~\ref{fig:epoch}(a), retrieval typically requires more epochs to converge, while ranking performance may decline with additional epochs. Achieving optimal performance in both stages is challenging without addressing these inconsistencies. Thus, \textbf{effectively monitoring and dynamically adjusting the training process across stages to achieve simultaneous performance improvements is crucial.}

To address these challenges, we propose the \textbf{Uni}fied \textbf{G}enerative \textbf{R}ecommendation \textbf{F}ramework (\textbf{UniGRF}), which unifies retrieval and ranking stages within a single generative model. We achieve this by converting tasks in both stages into sequence generation tasks and integrating them into a generative model by distinguishing outputs based on their positions. This approach enables sufficient information sharing between stages without additional costs and is model-agnostic and plug-and-play.
Then, to facilitate efficient collaboration between stages, we introduce a ranking-driven enhancer module within UniGRF. This module utilizes the high precision of the ranking stage to generate high-quality samples for the retrieval stage, creating an enhancement loop through parameter sharing.
Besides, to synchronize optimization across stages, UniGRF incorporates a gradient-guided adaptive weightier. This component dynamically monitors optimization rates based on gradients and adaptively weights the losses of different stages to balance optimization speeds. This allows the model to achieve optimal performance for both stages, fully leveraging the generative model's capabilities through a unified framework and efficient collaboration mechanism.
Finally, by optimizing the adaptive weighted loss, UniGRF enables end-to-end optimization of both stages, significantly enhancing their performance within a unified generative framework. The main contributions of this paper are as follows: 
\begin{itemize}[leftmargin=*,align=left]
    \item[$\bullet$] We creatively explore the construction of a Unified Generative Recommendation Framework, which focuses on efficient collaboration and synchronized optimization between retrieval and ranking stages within a single generative model. To our knowledge, this is the first framework to unify both stages in the generative recommendation. 
    \item[$\bullet$] We develop a ranking-driven enhancer module to facilitate inter-stage collaboration. This module efficiently generates high-quality samples with minimal computational overhead, creating a mutually enhancing loop between retrieval and ranking stages.
    \item[$\bullet$] We introduce a simple yet effective gradient-guided adaptive weightier that monitors optimization speeds of both stages, dynamically adjusting learning weights to achieve synchronous updates. This enables optimal performance of the generative model across both stages.
    \item[$\bullet$] Extensive experiments demonstrate that UniGRF outperforms state-of-the-art baselines on various benchmark datasets, highlighting the effectiveness of the unified framework. Ablation studies and further analysis confirm the UniGRF's strong scalability and its ability to achieve multi-stage collaboration and synchronized optimization in the generative recommendation.
\end{itemize}

\section{Related Works}
\subsection{Retrieval Models for Recommendation}
The retrieval stage efficiently selects a small subset of relevant candidates from all available items for subsequent processing~\cite{huang2024comprehensive,yin2024dataset,zhang2025td3,wang2025mf,wang2019mcne,wang2021hypersorec,wang2021decoupled,tong2024mdap,zhang2024unified,yin2024learning,xie2024breaking,yin2023apgl4sr,wang2024denoising,han2024efficient,han2023guesr}. 

Matrix Factorization (MF)~\cite{MatrixFactorization} is a foundational approach, representing users and items as vectors whose inner product indicates compatibility.
A significant advancement in this area is \emph{sequential recommendation}, where language modeling techniques shifted user representation to an autoregressive paradigm. Sequential models view users through the sequence of items they have interacted with. GRU4Rec \cite{gru4rec} applied recurrent neural networks to user behavior sequences, while BERT4Rec \cite{sun2019bert4rec} and SASRec \cite{sasrec} utilized transformers and self-attention for user behavior modeling.

Recently, large language models (LLMs) have influenced retrieval by enhancing user representation modeling~\cite{wu2024survey,wang2025generative,guo2024scaling,liu2023user,shen2024optimizing}. This is achieved through internal LLM representations and augmenting training data with LLM reasoning capabilities. For example, RLMRec \cite{ren2024representation} leverages LLMs for semantic representation to improve item vector quality, while LLMRec \cite{wei2024llmrec} uses LLMs to augment user-item interaction data alongside internal semantic representations.

LLMs have also inspired the first \emph{generative models} in the recommendation \cite{ye2025fuxi}. Meta's Hierarchical Sequential Transduction Units (HSTU) \cite{zhai2024actions} demonstrated that model capabilities scale with parameters, suggesting that traditional ranking and retrieval tasks can be reformulated as causal autoregressive problems, transforming them into sequence-to-sequence generation tasks. However, HSTU still trains models separately for the two stages. In contrast, ByteDance's Hierarchical Large Language Model (HLLM) \cite{chen2024hllm} incorporates a servant Item-LLM to extract textual information for item tokens, providing insights into integrating additional features into generative recommendation frameworks.

\vspace{-10pt}
\subsection{Ranking Models for Recommendation}
The primary goal of ranking models in recommendation systems is to reorder a refined pool of candidate items based on their relevance to the user, often using click-through rate (CTR) prediction~\cite{karatzoglou2013learning,wang2025universal,xu2024multi}. Early work focused on feature interaction, as seen in models like DCN \cite{wang2021dcn}, DeepFM \cite{Guo_2017}, AutoInt \cite{Song_2019}, and GDCN \cite{wang2023deeper}. The shift towards sequential recommendation has led to the adoption of autoregressive paradigms in ranking models, becoming crucial in industrial settings. The Deep Interest Network (DIN) \cite{zhou2018deep} introduced a \emph{target-aware} attention mechanism that dynamically reweights user behavior sequences using the candidate item vector as a query. DIEN \cite{zhou2018deepdien} enhanced DIN by incorporating GRU into the user's sequence, while SIM \cite{Pi_2020} used search techniques on long sequences to reduce noise in user sequence representations.

Similar to retrieval models, large language models (LLMs) have advanced ranking models~\cite{wu2024survey}. For instance, BAHE \cite{geng2024bahe} uses initial transformer blocks of an LLM to generate improved user behavior representations, with the final blocks fine-tuned for CTR prediction. LLMs have also driven the exploration of scaling laws in ranking models; generative recommendation models like HSTU \cite{zhai2024actions} have been adapted for ranking. The authors of HSTU suggest modifying it for ranking by appending candidate items to user sequences and adding MLP and prediction heads over these outputs. This formulation of target-aware attention aligns with how ranking is approached in generative recommendation settings, linking sequential ranking models with the generative recommendation.

\vspace{-10pt}
\subsection{Cascade Framework for Retrievers and Rankers}
As recommender systems grow more complex with multiple cascading stages (retrieval, pre-rank, rank, rerank), bias can propagate through the recommendation pipeline due to information loss in the multi-stage workflow. Consequently, research has focused on improving information transfer across stages to reduce sample bias.

Early methods~\cite{cascadeloss2} aimed to connect stage-specific objectives and jointly optimize them to balance efficiency and effectiveness globally. ECM~\cite{song2022rethinking} incorporates unexposed data into the training set and refines the task into a multi-label classification, addressing sample selection bias. RankFlow~\cite{qin2022rankflow} trains models for different stages separately and then jointly optimizes them by minimizing a global loss function to facilitate inter-stage information transfer. CoRR~\cite{corr} uses Kullback-Leibler (KL) divergence to distill knowledge between stages, aligning them for effective knowledge transfer. A recent approach~\cite{zheng2024full} introduces the Generalized Probability Ranking Principle (GPRP) to estimate selection bias and develop a ranking model that aligns with downstream module biases, optimizing multi-stage information retrieval systems.

Although these methods have been successful, they focus on traditional recommendation models and do not address efficient collaboration and synchronized optimization in the emerging field of generative recommendation.

\section{Problem Definition}
Let $\mathcal{U}=\{u_1, u_2, \ldots, u_{|\mathcal{U}|}\}$ and $\mathcal{I}=\{i_1, i_2, \ldots, i_{|\mathcal{I}|}\}$ represent the sets of users and items, with $|\mathcal{U}|$ and $|\mathcal{I}|$ denoting their respective sizes. For any user $u \in \mathcal{U}$, the interaction history is $X_u = \{i_1, b_1, \ldots, i_k, b_k, \ldots, i_n, b_n\}$, arranged chronologically. Here, $i_k \in \mathcal{I}$ is the $k$-th interacted item, and $b_k \in \mathcal{B} = \{0, 1\}$ indicates the interaction type, where $b_k = 1$ denotes a click, and $b_k = 0$ indicates no click. The sequence length $n$ is fixed; if interactions exceed $n$, only the most recent $n$ are retained. If fewer, the sequence is padded with the [padding] token to reach $n$.
The objective of generative recommendation is to predict the next item $i_{k+1}$ that user $u$ will interact with, based on $X_u$.

In practice, the recommendation process is divided into multiple stages for efficiency, with retrieval and ranking being crucial. We propose a unified framework that performs both stages simultaneously using a generative model:

\textbf{Unified Generative Recommendation Framework.}
This framework integrates retrieval and ranking in a generative model $\mathcal{M}$. In the retrieval stage, $\mathcal{M}$ selects a candidate set from $\mathcal{I}$ based on $X_u$, efficiently narrowing down potential items for user interaction. In the ranking stage, $\mathcal{M}$ scores these candidates and recommends $i_{k+1}$ based on the scores.

\section{Methodology}

\begin{figure*}[htbp]
\centering
\includegraphics[width=0.7\textwidth]{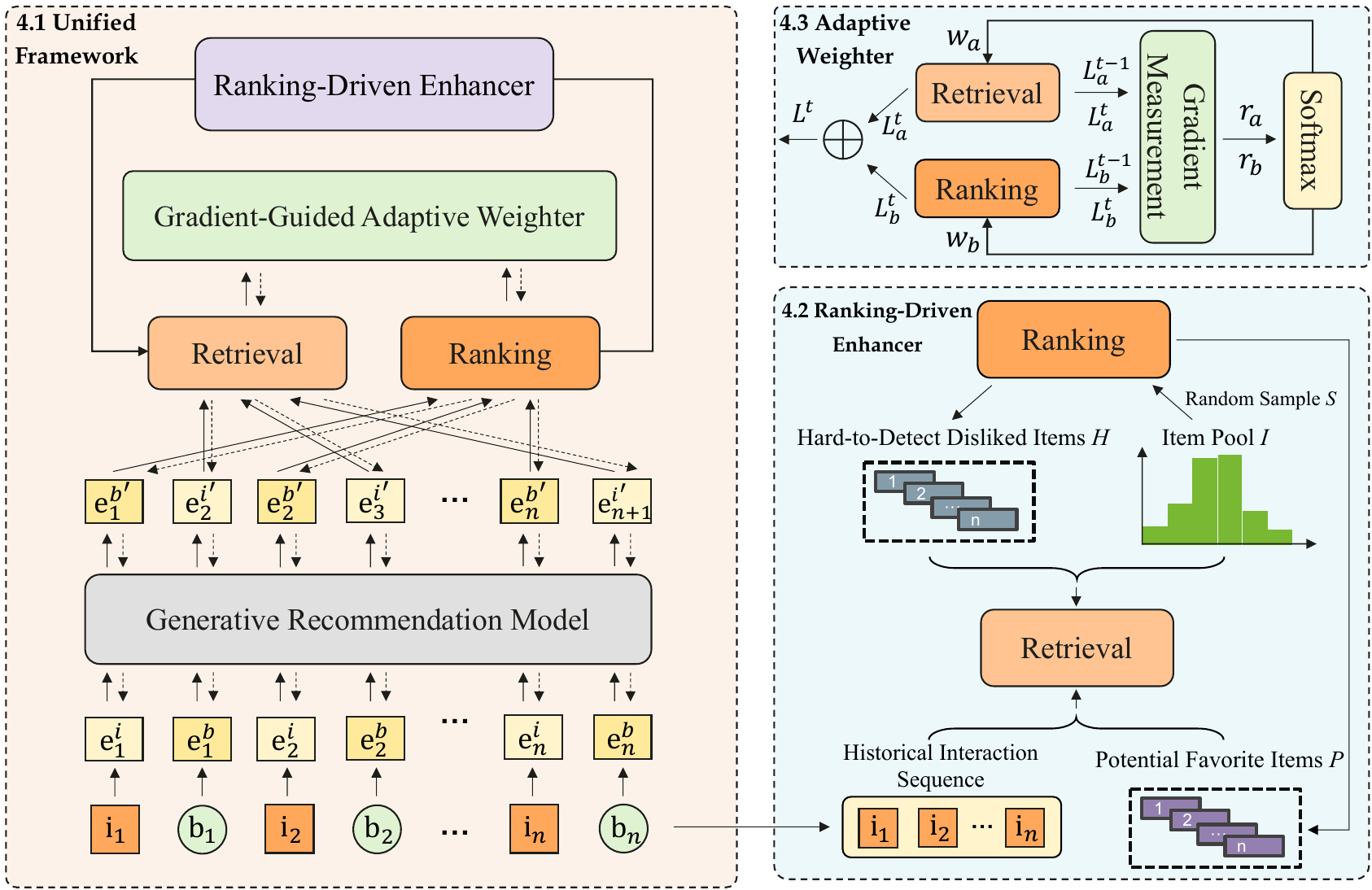}
\caption{Overview of UniGRF. This framework unifies retrieval and ranking into a generative recommendation model, with a ranking-driven enhancer and a gradient-guided adaptive weighter for efficient collaboration and synchronized optimization.} \label{overall}
\end{figure*}

In this section, we introduce the Unified Generative Recommendation Framework (UniGRF), which integrates the retrieval and ranking stages into a single generative model. An overview of UniGRF is shown in Fig.~\ref{overall}.
First, as detailed in Sec.~\ref{sec: framework}, UniGRF reformulates retrieval and ranking tasks as sequence generation tasks within a unified, model-agnostic framework. This approach ensures sufficient information transfer between stages and offers high scalability.
Second, in Sec.~\ref{sec: negative_sampler}, we introduce a ranking-driven enhancer module. This module generates accurate, high-quality samples, creating a mutually reinforcing loop with minimal computational overhead, thus facilitating efficient collaboration between stages.
Besides, Sec.~\ref{sec: weightedloss} describes a gradient-guided adaptive weighter. This component ensures synchronized performance improvement across stages, optimizing the overall framework.

\subsection{A Unified Generative Recommendation Framework}\label{sec: framework}

In recommendation systems, retrieval and ranking are crucial stages. Traditional generative methods often focus on a single stage or use separate models for each, leading to issues like data selection bias and information loss. To address these, we propose the Unified Generative Recommendation Framework (UniGRF), which integrates both stages within a single generative model.

For each user \( u \), items and interaction types in the user behavior sequence \( X_u = \{i_1, b_1, \ldots, i_k, b_k, \ldots, i_n, b_n\} \) are firstly encoded into embeddings, forming \( \mathbf{e}_u = \{\mathbf{e}_1^i, \mathbf{e}_1^b, \ldots, \mathbf{e}_k^i, \mathbf{e}_k^b, \ldots, \mathbf{e}_n^i, \mathbf{e}_n^b\} \). An autoregressive Transformer-based generative model refines these embeddings, producing outputs:
\begin{equation}
\text{Transformer}(\mathbf{e}_u) = \{\mathbf{e}_1^{b'}, \mathbf{e}_2^{i'}, \ldots, \mathbf{e}_k^{b'}, \mathbf{e}_{k+1}^{i'}, \ldots, \mathbf{e}_n^{b'}, \mathbf{e}_{n+1}^{i'}\},
\end{equation}
where \(\mathbf{e}_k^{b'}\) represents latent feedback for the \(k\)-th item, and \(\mathbf{e}_{k+1}^{i'}\) is the predicted next item.

The encoding process in our framework addresses two distinct learning objectives: first, predicting the click probability for the current item, represented by \(\mathbf{e}_k^{b'}\), which pertains to the ranking stage; second, forecasting the next item a user is likely to interact with, based on \(\mathbf{e}_{k+1}^{i'}\), which pertains to the retrieval stage. By clearly distinguishing these objectives at specific positions in the output, we effectively unify the retrieval and ranking stages within a single generative model. 

UniGRF is a plug-and-play generative recommendation framework that can be seamlessly integrated with any autoregressive generative model architecture, showcasing strong scalability. 
By unifying the retrieval and ranking stages into a single generative model, UniGRF effectively reduces both time and space complexity by half, regardless of the underlying architecture. This scalability and efficiency underscore its significant application potential. To ensure the model outputs align with our design, we introduce two constraints tailored to this framework.

\subsubsection{Constraint on the Retrieval Stage in UniGRF}
During the retrieval stage, the model predicts the next item \( \mathbf{e}_{k+1}^{i'} \) based on the user's previous \( k \) interactions. We assess the similarity of this predicted item with all items in the pool \( \mathcal{I} \) to identify the most similar item, which is then added to the candidate set \( \mathcal{I}^c \). This candidate set is subsequently refined in the ranking stage to achieve a more precise ordering.
 
To constrain this stage, we employ Sampled Softmax Loss~\cite{sampledsoftmaxloss} to optimize the model. Specifically, we aim to minimize the loss function such that the predicted item embedding vector \(\mathbf{e}_{k+1}^{i'}\) is as close as possible to the embedding vector of the next item that the user actually interacts with, \(\mathbf{e}_k^i\), in the vector space. Meanwhile, we randomly sample a set of negative samples \(\mathcal{S}\) from the item pool \(\mathcal{I}\), and we want \(\mathbf{e}_{k+1}^{i'}\) to be as distant as possible from the embeddings of these negative samples in the vector space. Specifically, the loss in the retrieval stage is as follows:
\begin{align*}
&\mathcal{L}_{\text{retrieval}} = \sum_{u \in \mathcal{U}} \sum_{k=2}^{n} 
\text{SampledSoftmaxLoss}(\mathbf{e}_k^{i'}) \\
&= -\sum_{u \in \mathcal{U}} \sum_{k=2}^{n}\log\left(\frac{\exp(\text{sim}(\mathbf{e}_k^{i'}, \mathbf{e}_k^i))}
{\exp(\text{sim}(\mathbf{e}_k^{i'}, \mathbf{e}_k^i)) + \sum_{j \in \mathcal{S}} \exp(\text{sim}(\mathbf{e}_k^{i'}, \mathbf{e}_k^j))}\right),
\end{align*}
where similarity $\text{sim}(\mathbf{a}, \mathbf{b}) = \mathbf{a} \cdot \mathbf{b}$, and $\cdot$ is the inner product.

\subsubsection{Constraint on the Ranking Stage in UniGRF}
The ranking stage is designed to assess user preferences by predicting the probability of a user clicking on items in the candidate set \( \mathcal{I}^c \). Items are then ranked based on these probability values, and items with a higher likelihood of being preferred are recommended to the user.

To predict this probability, we additionally introduce a small neural network. Specifically, for each input item \( i_k \), the model encodes the item into an intermediate vector \( \mathbf{e}_k^{b'} \). 
We then apply a small neural network \( f_{\phi}(\cdot) \) upon it to obtain the predicted click score \( score_k = \text{sigmoid}(f_{\phi}(\mathbf{e}_k^{b'})) \). 
This score is compared with the user's actual feedback \( b_k \) using binary cross-entropy loss:

\begin{equation}
\mathcal{L}_{ranking}= \sum_{u \in \mathcal{U}} \sum_{k=1}^{n} \text{BCELoss}(score_k, b_k).
\end{equation}

By unifying the retrieval and ranking stages into a single generative model, we address the issues of information loss and data bias inherent in previous generative recommendation methods.
Specifically, our approach draws negative samples from the entire item pool during the retrieval stage, allowing the model to learn the true data distribution and thereby reducing bias in the ranking stage. The use of a single generative model for both stages ensures that shared parameters facilitate an understanding of the data distribution across stages. 
Furthermore, the ranking stage captures more precise user interests, which enhances the performance of the retrieval task through knowledge transfer via shared parameters.

\subsection{Ranking-Driven Enhancer}\label{sec: negative_sampler}

In UniGRF, the retrieval and ranking stages are unified within a single generative model, allowing for simultaneous training. Despite this integration, the model currently lacks a collaborative mechanism to effectively leverage information from both stages for mutual enhancement. To address this, we introduce a ranking-driven enhancer module. This module promotes inter-stage collaboration and improves user preference modeling by accurately identifying potential favorite items and hard-to-detect disliked dislikes.

\subsubsection{Hard-to-Detect Disliked Items Generator}

In retrieval stages, negative samples \(\mathcal{S}\) are randomly selected from the item pool \(\mathcal{I}\) due to computational constraints. This approach often results in negative samples that are too easily distinguishable from positive ones, thereby limiting the model's training effectiveness. To address this, we utilize the ranking model's precision to identify more challenging negative samples, termed hard-to-detect disliked items, denoted as \(\mathcal{H}\). These samples contain richer information and enhance the retrieval stage's training. By repeatedly training the model on \(\mathcal{H}\) until it can effectively distinguish these samples, we facilitate information transfer from the ranking stage to the retrieval stage.

To construct a hard-to-detect disliked item set $\mathcal{H}$, we analyze the relative scores from both the retrieval and ranking stages. Items that receive high scores in the retrieval stage but low scores in the ranking stage are identified as hard-to-detect disliked items. 

The ranking stage score for item in negative samples \(\mathcal{S}\) is calculated as follows:
\begin{equation}
{score}_{\text{ranking}} = \text{sigmoid}(f_{\phi}(\mathbf{e}^\mathcal{S})),
\end{equation}
where $\mathbf{e}^\mathcal{S}$ represents embedding of the negative sample set $\mathcal{S}$ and \( f_{\phi}(\cdot) \) is the scoring function in ranking stage.

Concurrently, the retrieval score ${score}_{\text{retrieval}}$ is calculated as:

\begin{equation}
{score}_{\text{retrieval}} = \text{sigmoid}(\text{sim}(\mathbf{e}^{i'}, \mathbf{e}^\mathcal{S})),
\end{equation}
where $\mathbf{e}^{i'}$ is the predicted user-interacted item embedding.

Then, the relative score is defined as:

\begin{equation}
{score} = {score}_{\text{retrieval}} \cdot \left(\frac{{score}_{\text{retrieval}}}{{score}_{\text{ranking}}} - 1\right).
\end{equation}

We select the top \(m\) items with the highest relative scores as hard-to-detect dislikes and include them in \(\mathcal{H}\). These hard negatives are then combined with randomly sampled negatives to form a new negative sample set \(\mathcal{S}\) for subsequent training epoch. This process continues until the model can effectively distinguish these items, at which point they receive low scores and are removed from \(\mathcal{H}\). By continuously integrating these hard-to-detect disliked items into the training process, we introduce more informative data to the retrieval stage, thereby enhancing the model's training efficiency.

\subsubsection{Potential Favorite Items Generator}
In the negative sample set $\mathcal{S}$, there is a possibility that some items might actually be liked by the user, which can lead to inaccuracies when these items are randomly selected and labeled as negative. This mislabeling can adversely affect the model's performance by introducing bias into the training process. To address this issue, we generate potential favorite items for the user based on the ranking stage's precise understanding of user intent, providing more accurate training samples for the retrieval stage.

The module scores each item in the sample set $\mathcal{S}$, which comes from the previous epoch and is initially generated through random negative sampling.
Items that receive higher scores from the ranking model are identified as potentially liked by the user. Specifically, we define a threshold, $\alpha$, to refine this selection process: items with a ranking model score ${score}_{\text{ranking}} > \alpha$ are flagged as potential liked items and are subsequently stored in the set $\mathcal{P}$.

For potentially liked items in $\mathcal{P}$, incorrectly treating them as negative samples can mislead the generative model's learning process. To mitigate this, we reassign their labels to positive in the subsequent iteration. This adjustment helps the generative model accurately learn user preferences by incorporating these items as positive examples, thereby correcting any previous misclassification and enhancing overall model performance.

Through the Hard-to-Detect Disliked Items Generator and the Potential Favorite Items Generator, UniGRF leverages the ranking stage's more precise understanding of user intent to generate high-quality training data for the retrieval stage, while maintaining the original data distribution. Notably, this guidance is not one-way. Since the recommendation tasks of both stages are unified within a model with fully shared parameters, improvements in the retrieval stage's understanding of data distribution will also benefit the ranking stage, creating a mutually reinforcing feedback loop. Notably, the ranking score is calculated through a linear complexity function $f_{\phi}(\cdot)$, while the retrieval score calculation is integrated into the retrieval loss. This design ensures that the additional computational overhead of the module is minimal. Thus, the ranking-driven enhancer module introduces an efficient and low-cost collaborative mechanism for UniGRF, achieving bidirectional enhancement by leveraging the interdependence between both stages.
 
\subsection{Gradient-Guided Adaptive Weighter}\label{sec: weightedloss}

Following the collaborative enhancement of the retrieval and ranking stages via the Ranking-Driven Enhancer, a new challenge arises within the fully-shared unified framework: the retrieval and ranking stages exhibit significantly different convergence speeds, complicating simultaneous optimization. To address this issue, we introduce a Gradient-Guided Adaptive Weighter. This mechanism evaluates the learning speeds of the retrieval and ranking stages by monitoring their gradient update rates at various time steps. It dynamically adjusts the weights of these stages to align their convergence speeds, thereby facilitating synchronous optimization and enhancing overall performance.

Specifically, the model updates parameters by back-propagating gradients. The faster the loss value decreases, the larger the gradient changes, and the faster the task is updated. Therefore, we use the magnitude of the loss decrease to qualitatively analyze the convergence speed:

\begin{align}
r_{a} = \mathcal{L}_{retrieval}^t/\mathcal{L}_{retrieval}^{t-1}, \\
r_{b} = \mathcal{L}_{ranking}^t/\mathcal{L}_{ranking}^{t-1},
\end{align}
where \(\mathcal{L}_{retrieval}^t\) and \(\mathcal{L}_{ranking}^t\) represent the retrieval and ranking losses at the \(t\)-th time step, respectively. The convergence rates \(r_{a}\) and \(r_{b}\) represent the relative optimization magnitude of the retrieval and ranking tasks. A larger \(r\) value indicates slower convergence, so a larger weight should be given to this task. This gradient-based optimization speed approximation method can dynamically monitor the update rates of different stages to adjust the optimization speed dynamically.

Based on this, we adaptively calculate the weights:

\begin{gather}
w_a = \frac{\lambda_a\text{exp}(r_a / T)}{\text{exp}(r_a / T) + \text{exp}(r_b / T)}, \\
w_b = \frac{\lambda_b\text{exp}(r_b / T)}{\text{exp}(r_a / T) + \text{exp}(r_b / T)},
\end{gather}
where \(T\) is the temperature coefficient, which is used to adjust the weight difference between retrieval and ranking. Smaller \(T\) values lead to larger weight differences. \(\lambda_a\) and \(\lambda_b\) are hyperparameters used to scale the losses of the two stages to the same magnitude.

After adaptively adjusting the weights, we then dynamically weight the retrieval and ranking losses to obtain the final loss function $\mathcal{L}^t$ for optimization:

\begin{equation}
\mathcal{L}^t = w_a \cdot \mathcal{L}_{retrieval}^t + w_b \cdot \mathcal{L}_{ranking}^t.
\end{equation}

This gradient-guided adaptive weighted loss \(\mathcal{L}^t\) enables tasks in the retrieval and ranking stages to be optimized synchronously. It allows the model to meet the needs of both stages simultaneously, thereby leveraging the sufficient information transfer provided by the unified framework and the efficient collaboration mechanism brought by the Ranking-Driven Enhancer, ultimately enhancing the recommendation performance of both stages.

\section{Experiments}

\subsection{Experiments Settings}

\subsubsection{Datasets}

\begin{table}[t]
\centering
\caption{Statistical information of experimental datasets.}
\begin{tabular}{lrrrrr}
\toprule
\textbf{Datasets} & \textbf{\#Users} & \textbf{\#Items} & \textbf{\#Inters.} & \textbf{Avg. $n$} \\
\midrule
MovieLens-1M & 6,040 & 3,706 & 1,000,209 & 165.6 \\
MovieLens-20M & 138,493 & 26,744 & 20,000,263 & 144.4 \\
Amazon-Books & 694,897 & 686,623 & 10,053,086 & 14.5 \\
\bottomrule
\end{tabular}

\label{tab:datasets}
\end{table}

To verify the effectiveness of our proposed unified generative framework, UniGRF, we conduct experiments on three public recommendation datasets of varying sizes: MovieLens-1M, MovieLens-20M, and Amazon-Books. These datasets are derived from real-world online interactions of real users and are widely used in the research of recommendation systems.

\begin{itemize}[left=0pt]
\item \textbf{MovieLens-1M\footnote{\url{https://grouplens.org/datasets/movielens/1m/}}} \cite{harper2015movielens}: This dataset contains approximately 1 million user rating records for movies, involving about 6,000 users and 3,900 movies. The ratings range from 1 to 5 and include basic information about users and metadata about movies.
\item \textbf{MovieLens-20M\footnote{\url{https://grouplens.org/datasets/movielens/20m/}}} \cite{harper2015movielens}: This larger dataset contains about 20 million rating records, covering approximately 138,000 users and 27,000 movies, spanning from 1995 to 2015.
\item \textbf{Amazon-Books\footnote{\url{http://snap.stanford.edu/data/amazon/productGraph/categoryFiles/ratings_Books.csv}}} \cite{mcauley2013hidden}: This dataset is a subset of Amazon product reviews, focusing on the book category. It contains millions of user ratings and reviews.
\end{itemize}

For data processing, we arrange each user's interaction records in chronological order to form a user historical interaction sequence and filtered out sequences with fewer than three interactions. For the rating data, we apply binary processing: ratings greater than 3 are marked as 1, indicating the user liked the item; otherwise, they are marked as 0. This method simplifies the representation of user interests. The details of the datasets are shown in Table~\ref{tab:datasets}.

\subsubsection{Compared Methods}
We compare our method with classic state-of-the-art retrieval and ranking methods, which are typically trained separately, and also consider a cascade framework that combines the relevance of retrieval and ranking. The specific comparison methods are as follows:

\begin{flushleft}
\textbf{Retrieval Models}:
\end{flushleft}
   \begin{itemize}[left=0pt]
        \item \textbf{BPR-MF}~\cite{rendle2012bpr}: BPR-MF is based on matrix factorization and Bayesian personalized ranking (BPR) framework. 
        \item \textbf{GRU4Rec}~\cite{jannach2017recurrent}: GRU4Rec uses recurrent neural networks (RNNs) and gated recurrent units (GRUs) for the recommendation.
        \item \textbf{NARM}~\cite{li2017neural}: NARM captures users' different preferences by combining recurrent neural networks and attention mechanisms.
        \item \textbf{SASRec}~\cite{sasrec}: SASRec captures complex patterns in user behaviors through the self-attention layer.
    \end{itemize}
\textbf{Ranking Models}:
\begin{itemize}[left=0pt]
   \item \textbf{DCN}~\cite{wang2021dcn}: DCN combines deep networks and cross networks to capture high-order interactions between features. 
   \item \textbf{DIN}~\cite{zhou2018deep}: DIN introduces an attention mechanism to dynamically model the user's interest distribution.
   \item \textbf{GDCN}~\cite{wang2023towards}: GDCN captures complex relationships and contextual information between features by propagating and aggregating information with a gated cross network.
\end{itemize}
\textbf{Generative Models}: We select two generative models whose architectures support retrieval and ranking tasks. Notably, these models are designed to be trained independently for a single task rather than performing both tasks simultaneously.
\begin{itemize}[left=0pt]
   \item \textbf{HSTU}~\cite{zhai2024actions}: HSTU significantly improves training and inference efficiency through the point-wise aggregated attention mechanism and the M-FALCON algorithm. 
   \item \textbf{Llama}~\cite{touvron2023llama}: Llama is originally designed as a Transformer-based architecture for natural language processing (NLP) tasks. We adopt this architecture to build a recommendation system and use the sequence generation method from HSTU to perform retrieval and ranking.
\end{itemize}
\textbf{Cascade Frameworks for Generative Model}:
These frameworks address inter-stage information transfer by typically using two separate traditional models with specialized modules for information exchange. We instantiate the retrieval and ranking modules in these frameworks as HSTU to verify their effectiveness in generative recommendation scenarios.

\begin{itemize}[left=0pt]
    \item \textbf{RankFlow-HSTU}~\cite{qin2022rankflow}: The framework first performs independent training and then joint training to achieve the transfer and sharing of information between the retrieval and ranking stages.
   \item \textbf{CoRR-HSTU}~\cite{corr}: The CoRR framework introduces a knowledge distillation method utilizing Kullback-Leibler (KL) divergence to align the retrieval and ranking stages effectively. 
\end{itemize}

\subsubsection{Evaluation Protocols}

To ensure a fair comparison, we employ the same training and test set partitioning strategy for all methods. Specifically, the item from the user's last interaction is used as the test set, the item from the second-to-last interaction is used as the validation set, and the items from all other interactions are used as the training set.
To avoid sampling bias in the candidate set, the entire item set is used as the candidate set for retrieval~\cite{usersampled}.
For evaluating retrieval performance, we use NDCG@K, HR@K, and MRR as metrics, which are widely used in related works~\cite{ndcg10,hr10,mrr}, and we set the K value to 10 and 50. We use AUC~\cite{zhou2018deep} to measure ranking performance.

\subsubsection{Parameter Settings}
To ensure a fair comparison, we use the same hyperparameter settings for the same dataset across all baseline models, with the learning rate uniformly set to $1e^{-3}$. For the retrieval stage, the MovieLens-1M dataset uses 128 negative samples, while the MovieLens-20M and Amazon-Books datasets use 256 samples. Regarding the number of training epochs, the ranking-only model trains for 20 epochs, with early stopping applied if the AUC value does not improve over 5 epochs. The other models train for 100 epochs, also using an early stopping strategy. 
For Transformer-based models, the number of layers is fixed at 2 for the MovieLens-1M dataset and at 4 for the MovieLens-20M and Amazon-Books datasets unless otherwise specified. All models are trained in a distributed manner, utilizing 8 cards for computation.

\subsection{Overall Performance}

\begin{table*}[t]
\centering
\caption{Comparison of retrieval performance across three datasets. The best result is highlighted in bold, and the second-best is underlined. (p-value < 0.05)}\label{overall_retrieval}
\renewcommand{\arraystretch}{1.3}
\resizebox{\textwidth}{!}{ 
\begin{tabular}{l|lllll|lllll|lllll}
\hhline{=|=====|=====|=====}
\textbf{Dataset}                       & \multicolumn{5}{c|}{\textbf{MovieLens-1M}}                                                                                                      & \multicolumn{5}{c|}{\textbf{MovieLens-20M}}         & \multicolumn{5}{c}{\textbf{Amazon-Books}}    \\ \hline
\textbf{Model} & \textbf{NG@10} & \textbf{NG@50} & \textbf{HR@10} & \textbf{HR@50} & \textbf{MRR}    & \textbf{NG@10}  & \textbf{NG@50}  & \textbf{HR@10}  & \textbf{HR@50}  & \textbf{MRR}    & \textbf{NG@10} & \textbf{NG@50} & \textbf{HR@10} & \textbf{HR@50} & \textbf{MRR} \\ \hline
BPRMF & 0.0607 & 0.1027 & 0.1185 & 0.3127 &  0.0556 & 0.0629 & 0.1074 & 0.1241 & 0.3300 & 0.0572 & 0.0081 & 0.0135 & 0.0162 & 0.0412 & 0.0078 \\
GRU4Rec & 0.1015 & 0.1460 & 0.1816 & 0.3864 & 0.0895 & 0.0768 & 0.1155 & 0.1394 & 0.3177 & 0.0689 & 0.0111 & 0.0179 & 0.0207 & 0.0519 & 0.0107  \\
NARM & 0.1350 & 0.1894 & 0.2445 & 0.4915 & 0.1165 & 0.1037 & 0.1552 & 0.1926 & 0.4281 & 0.0910 & 0.0148 & 0.0243 & 0.0280 & 0.0722 & 0.0141\\
SASRec                        & 0.1594     & 0.2208     & 0.2899     & 0.5671     & 0.1388      & 0.1692      & 0.2267      & 0.3016      & 0.5616      & 0.1440      & 0.0229     & 0.0353     & 0.0408     & 0.1059    & 0.0213   \\ \hline
Llama                         & 0.1746     & \underline{0.2343}     & 0.3110     & 0.5785     & 0.1477      & \underline{0.1880}      & \underline{0.2451}      & \underline{0.3268}      & \textbf{0.5846}      & \underline{0.1605}      & 0.0338     & 0.0493     & 0.0603     & 0.1317    & 0.0305   \\
HSTU                          & 0.1708 & 0.2314 & 0.3132 & 0.5862 & 0.1431 & 0.1814 & 0.2325 & 0.3056 & 0.5371 & 0.1569 & 0.0337     & 0.0502     & 0.0613     & \underline{0.1371}     & 0.0305   \\ \hline
RankFlow-HSTU                         & 0.1096     & 0.2178      & 0.2184     & 0.5062    & 0.0988      & 0.0985      & 0.2064      & 0.2002      & 0.4245       & 0.1221     & 0.0281   & 0.0307     & 0.0324     & 0.0878      & 0.0249   \\
CoRR-HSTU                         & 0.1468     & 0.2662      & 0.2081     & 0.5435    & 0.1261      & 0.1629      & 0.2166      & 0.2796      & 0.5236       & 0.1409     & \underline{0.0352}   & \underline{0.0521}     & 0.0612     & 0.1360      & 0.0315 \\ \hline
\textbf{UniGRF-Llama}                       & \textbf{0.1765}     & \textbf{0.2368}     & \textbf{0.3219}     & \textbf{0.5921}     & \underline{0.1478}      & \textbf{0.1891}      & \textbf{0.2464}      & \textbf{0.3270}      & \textbf{0.5846}      & \textbf{0.1652}      & 0.0351     & 0.0508     & \underline{0.0624}     & 0.1347     & \underline{0.0316}  \\
\textbf{UniGRF-HSTU}                        & \underline{0.1756} & 0.2326 & \underline{0.3140} & \underline{0.5909} & \textbf{0.1484} & 0.1816 & 0.2384 & 0.3178 & \underline{0.5747} & 0.1548 & \textbf{0.0354}     & \textbf{0.0522}     & \textbf{0.0638}     & \textbf{0.1415}     & \textbf{0.0319}   \\ \hhline{=|=====|=====|=====}
\end{tabular}
}
\end{table*}

\begin{table}[t]
\centering
\caption{Comparison of ranking performance across three datasets. The best result is highlighted in bold, and the second-best is underlined. (p-value < 0.05)}\label{overall_ranking}
\renewcommand{\arraystretch}{0.9}
\resizebox{0.7\columnwidth}{!}{ 
\begin{tabular}{l|c|c|c}
\hhline{=|=|=|=}
\textbf{Dataset} & \multicolumn{1}{c|}{\textbf{ML-1M}} & \multicolumn{1}{c|}{\textbf{ML-20M}} & \multicolumn{1}{c}{\textbf{Books}} \\ \hline
\textbf{Model} & \textbf{AUC} & \textbf{AUC} & \textbf{AUC} \\ \hline
DCN & 0.7176 & 0.7098 & 0.7042 \\
DIN & 0.7329 & 0.7274 & 0.7159 \\
GDCN & 0.7364 & 0.7089 & 0.7061 \\ \hline
Llama & 0.7819 & 0.7722 & 0.7532 \\
HSTU & 0.7621 & \underline{0.7807} & 0.7340 \\ \hline
RankFlow-HSTU & 0.6340 & 0.6439 & 0.6388 \\
CoRR-HSTU & 0.7497 & 0.7492 & 0.7357 \\ \hline
\textbf{UniGRF-Llama} & \textbf{0.7932} & 0.7776 & \underline{0.7559} \\
\textbf{UniGRF-HSTU} & \underline{0.7832} & \textbf{0.7941} & \textbf{0.7672} \\ \hhline{=|=|=|=}
\end{tabular}
}

\end{table}
 
To ensure a fair and comprehensive comparison, we instantiate UniGRF using the representative generative models HSTU and Llama, denoting them as UniGRF-HSTU and UniGRF-Llama, respectively. Table \ref{overall_retrieval} and Table \ref{overall_ranking} display the retrieval and ranking results of the baseline methods and UniGRF across three datasets. From this performance comparison, we can draw the following conclusions:

\begin{itemize}[left=0pt]
    \item In scenarios where different models are designed for retrieval and ranking, generative methods such as HSTU and Llama consistently outperform non-generative methods in recommendation performance. This suggests that generative models possess significant potential for development by learning richer feature representations and deeper user interest relationships.

    \item Our framework, UniGRF, consistently outperforms all baselines in both the retrieval and ranking stages, whether based on the HSTU or Llama architecture. This demonstrates that UniGRF effectively enhances performance through robust information transfer between stages, achieving efficient inter-stage collaboration and synchronous optimization. Additionally, UniGRF is model-agnostic and offers a promising unified framework for generative recommendation. In contrast, cascade frameworks like RankFlow-HSTU and CoRR-HSTU perform poorly and even negatively impact performance. This suggests that traditional cascade frameworks fail to adapt to generative recommendation models and cannot facilitate the necessary inter-stage information transfer to boost performance.
    \item While our framework achieves improvements in the retrieval stage, it demonstrates a more significant performance boost in the ranking stage. This is likely because the retrieval stage benefits from learning through sampling negative examples, whereas the ranking stage, with its simpler task, lacks sufficient information on its own. Consequently, training the ranking module separately results in relatively poorer performance, making it more amenable to enhancement through inter-stage information transfer. This trend positively impacts the practical application of UniGRF. The retrieval stage primarily generates a large candidate set, which has a smaller direct impact on the final results and is less sensitive to metric changes. In contrast, the ranking stage directly influences the quality of the recommendation results. Therefore, UniGRF's superior performance in the ranking stage is particularly crucial, as it can significantly enhance overall recommendation performance.
    \item On the Amazon-Books dataset, the UniGRF model shows a more significant performance improvement compared to other datasets. Additionally, the CoRR cascade framework also performs relatively well on this dataset. We speculate that this phenomenon may be related to the dataset's sparsity. Sparse data limits the availability of sufficient information, making cross-stage information transfer particularly crucial in this context.
\end{itemize}

\subsection{Ablation Study}
To verify the effectiveness of each module, an ablation experiment is conducted to compare our framework UniGRF with its three variants: (1). \textit{\textbf{w/o Enhancer}}: removes the Ranking-Driven Enhancer module and uses random sampling to generate the same number of negative samples; (2). \textit{\textbf{w/o Weighter}}: removes the Gradient-Guided Adaptive Weighter and simply adds the retrieval and ranking losses as the optimization target, i.e., \(\mathcal{L}^t = \mathcal{L}_{retrieval}^t + \mathcal{L}_{ranking}^t\); (3). \textit{\textbf{w/o Both}}: removes both the Ranking-Driven Enhancer module and the Gradient-Guided Adaptive Weighter. To facilitate comparison, all recommendation modules within the framework are instantiated using the HSTU model.

In addition to performing ablation experiments on each module of the framework, we further verify the effectiveness of information transfer in the generative recommendation by comparing our unified framework, UniGRF, with the fine-tuning paradigm. Specifically, we design the following two experimental modules: (1). \textit{\textbf{HSTU-FT-a}}: The HSTU model is first used to instantiate the retrieval model, which is trained for 100 epochs. The trained parameters are then used to initialize the ranking model, which is subsequently trained for 20 epochs. The fine-tuned model is applied in the ranking stage; (2). \textit{\textbf{HSTU-FT-b}}: The HSTU model is initially used to instantiate the ranking model, which is trained for 20 epochs. The retrieval model is then initialized using the parameters from the ranking model, and training is continued for 100 epochs. The fine-tuned model is applied in the retrieval stage.
The results on the MovieLens-1M dataset of the ablation experiment are listed in Table~\ref{ablation}. To more conveniently present the results, we also include the results of the separately trained model (denoted as HSTU) and our unified generative framework (UniGRF-HSTU).
\begin{table}[t]
\centering
\caption{Ablation analysis on MovieLens-1M.}\label{ablation}
\renewcommand{\arraystretch}{1.3}
\resizebox{\columnwidth}{!}{ 
\begin{tabular}{l|l|lllll}
\hhline{=|=|=====}
\textbf{Model} & \textbf{AUC} & \textbf{NG@10} & \textbf{NG@50} & \textbf{HR@10} & \textbf{HR@50} & \textbf{MRR} \\ \hline
HSTU & 0.7621 & 0.1708 & 0.2314 & 0.3132 & \underline{0.5862} & 0.1431 \\ \hline
(1) w/o Enhancer & \underline{0.7815} & 0.1717 & \underline{0.2318} & 0.3117 & 0.5824 & \underline{0.1446} \\
(2) w/o Weighter & 0.7582 & \underline{0.1734} & 0.2317 & \textbf{0.3199} & 0.5824 & 0.1441 \\
(3) w/o Both & 0.7293 & 0.1681 & 0.2276 & 0.3113 & 0.5824 & 0.1392 \\\hline
(a) HSTU-FT-a & 0.7752 & - & - & - & - & - \\
(b) HSTU-FT-b & - & 0.1727 & 0.2308 & \underline{0.3182} & 0.5807 & 0.1436 \\ \hline
\textbf{UniGRF-HSTU} & \textbf{0.7832} & \textbf{0.1756} & \textbf{0.2326} & 0.3140 & \textbf{0.5909} & \textbf{0.1484} \\ \hhline{=|=|=====}
\end{tabular}
}
\end{table}

The results in Table~\ref{ablation} indicate that the performance of all variants is inferior to UniGRF-HSTU, demonstrating the effectiveness of the proposed Ranking-Driven Enhancer module and Gradient-Guided Adaptive Weighter. The specific analysis is as follows: 
\begin{itemize}[left=0pt]
    \item The variant "w/o Both" exhibits the worst performance, failing to surpass the independently optimized "HSTU" model. This outcome verifies our insights that simply unifying retrieval and ranking into one generative framework is insufficient for performance improvement. Instead, neglecting the collaboration mechanism between stages and lack of synchronized optimization can lead to performance degradation. Therefore, exploring how to achieve such a promising unified framework for generative recommendation is necessary. 
    \item In the "w/o Enhancer" variant, the use of uniform negative sampling leads to performance degradation in the retrieval stage. This suggests that, unlike the Ranking-Driven Enhancer, uniform negative sampling fails to capture users' deep interests and may introduce noise, reducing the generative model's effectiveness. In comparison, the "w/o Weighter" variant shows significant degradation in ranking performance compared to UniGRF-HSTU due to the absence of an adaptive weight adjustment mechanism. This highlights the critical role of the Gradient-Guided Weighter in achieving synchronized optimization and reaching optimal performance across both stages. However, the "w/o Weighter" variant still performs better than the "w/o Both" variant. This indicates that the Ranking-Driven Enhancer can improve generative recommendation model performance by enhancing inter-stage collaboration. Additionally, this improvement may stem from the module's ability to focus the model on samples selected during the ranking stage, thereby potentially enhancing the supervision signal in the ranking stage.
    \item Compared to the base model HSTU, the model based on fine-tuned paradigms, HSTU-FT-a and HSTU-FT-b, demonstrate improved performance. This indicates that transferring information between the retrieval and ranking stages through a specific mechanism is effective. However, despite more training epochs, the performance of HSTU-FT-a and HSTU-FT-b does not surpass our framework UniGRF-HSTU. This suggests that employing the unified generative framework for both retrieval and ranking stages facilitates more sufficient information transfer.
\end{itemize}
\begin{figure}[t]
    \centering
    \begin{subfigure}{0.236\textwidth}
        \centering
        \includegraphics[width=\linewidth]{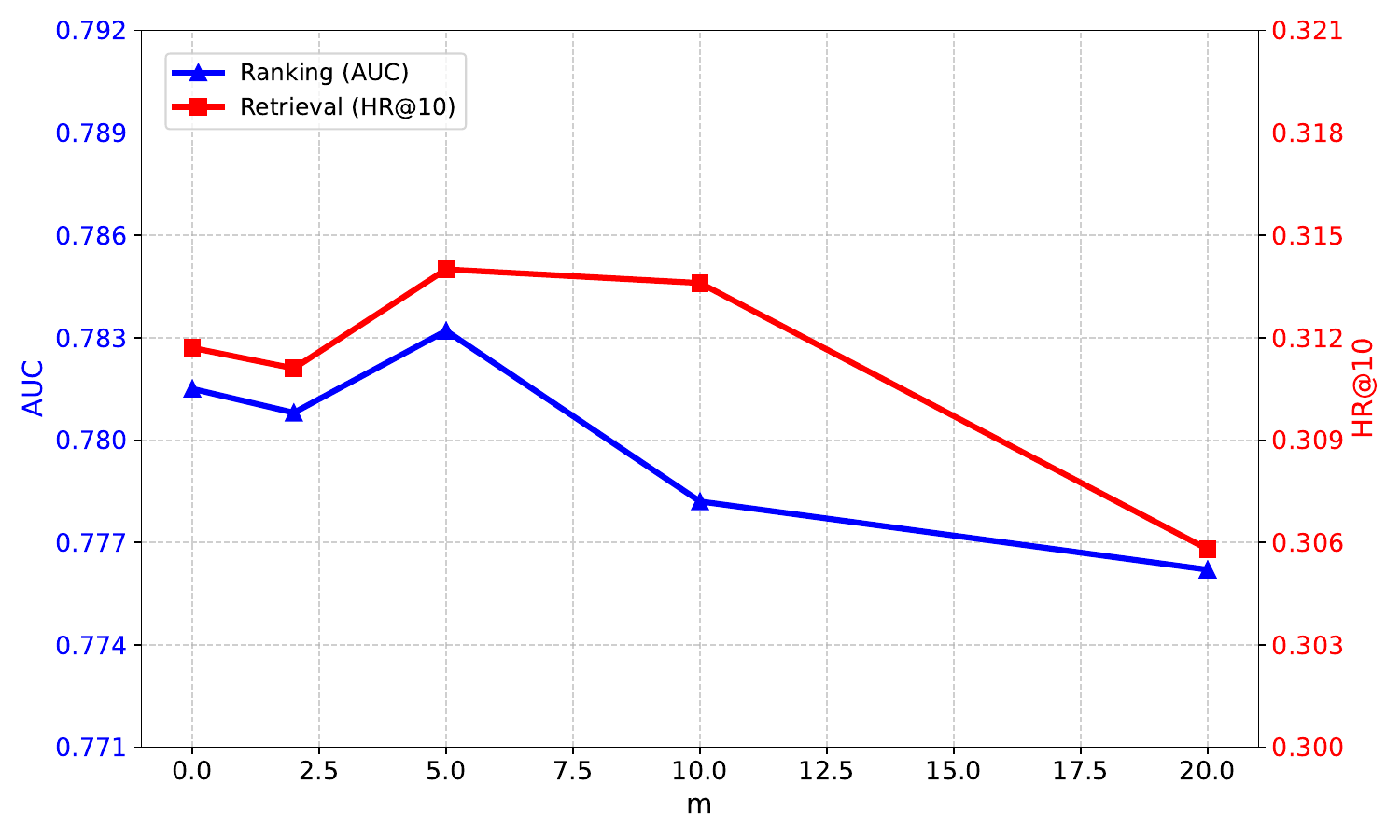}
        \caption{MovieLens-1M}
        \label{fig:subfig1numsample}
    \end{subfigure}
    \hfill
    \begin{subfigure}{0.236\textwidth}
        \centering
        \includegraphics[width=\linewidth]{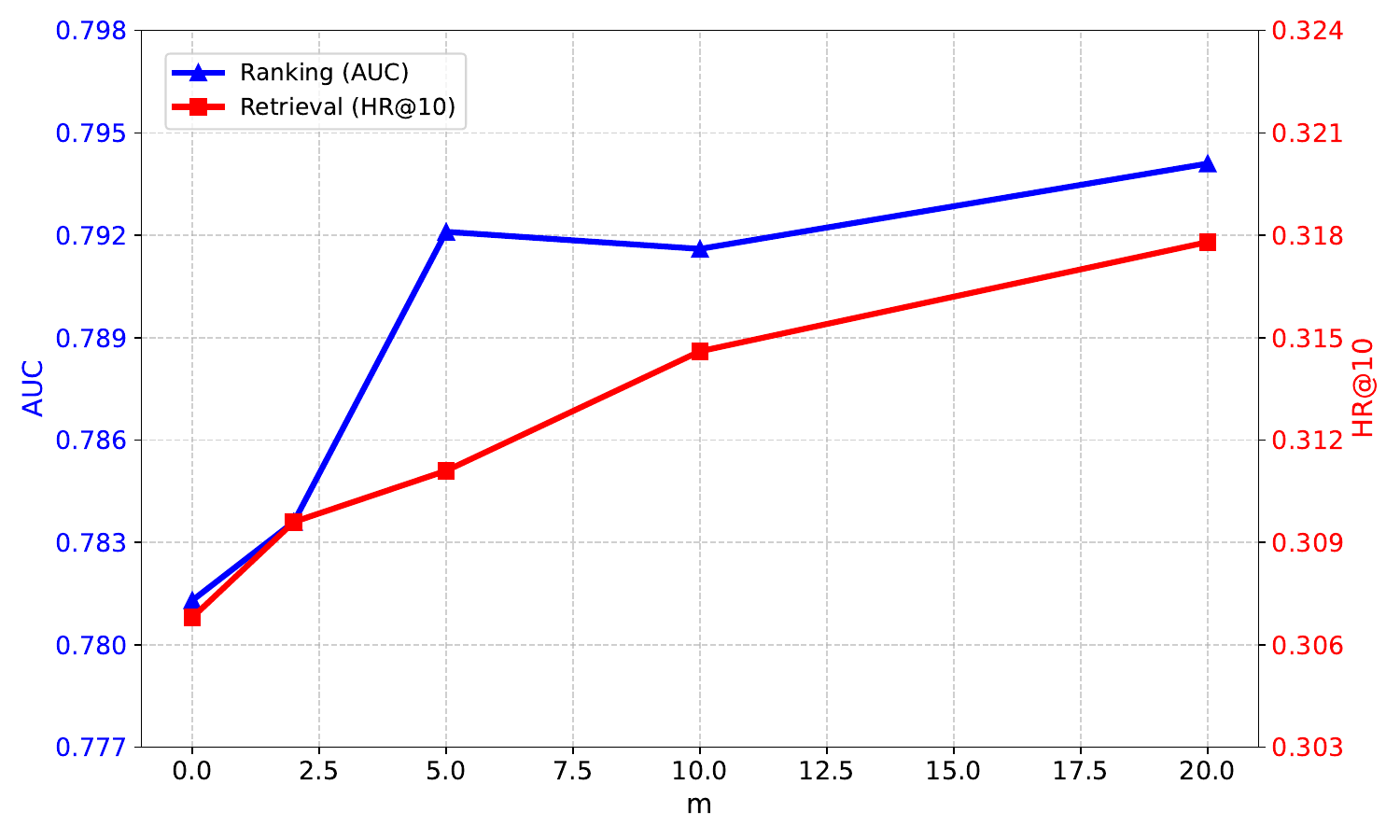}
        \caption{MovieLens-20M}
        \label{fig:subfig2numsample}
    \end{subfigure}
    \caption{Impact of varying the number of hard-to-detect disliked items $m$.}
    \label{fig:numsample}
\end{figure}

\vspace{5pt}
\subsection{Impact of Hyper-parameter}

UniGRF generates $m$ items containing the user's fine-grained negative interests through the Ranking-Driven Enhancer. To investigate the effect of introducing more challenging samples, we fix the total number of negative samples (128 for MovieLens-1M and 256 for MovieLens-20M) and vary the value of the number of hard-to-detect disliked items $m$ to $\{0, 2, 5, 10, 20\}$. Fig.~\ref{fig:numsample} presents the results on the MovieLens-1M and MovieLens-20M datasets, which are the smallest and largest datasets in this paper, respectively. 

For the larger dataset MovieLens-20M, as $m$ increases from 0 to 20, the performance of both the retrieval and ranking stages shows a continuous upward trend. This indicates that introducing difficult negative samples through the Ranking-Driven Enhancer can facilitate the model in learning the user's accurate preferences and achieve mutual enhancement between stages, thereby enhancing recommendation performance. 

For the smaller MovieLens-1M dataset, although the performance of both the retrieval and ranking stages initially increases and then decreases as $m$ increases from 0 to 20, it reaches an optimal value around $m=5$. This indicates that in smaller datasets, moderately introducing hard negative samples can effectively enhance the model's learning ability, helping it better capture users' fine-grained preferences. However, when the value of $m$ is too large, the model may over-focus on these hard negative samples, leading to overfitting and thus affecting overall performance. Additionally, when the proportion of hard negative samples is too high, the model may repeatedly encounter the same negative samples in each training epoch. This overlap reduces the diversity of training samples, causing the model to lack sufficient variation and fresh information during the learning process, thereby limiting its generalization ability. Therefore, on small datasets, selecting an appropriate number of hard negative samples is crucial to balance improving model performance and avoiding overfitting.

\begin{figure}[tbp]
    \centering
    \begin{subfigure}{0.236\textwidth}
        \centering
        \includegraphics[width=\linewidth]{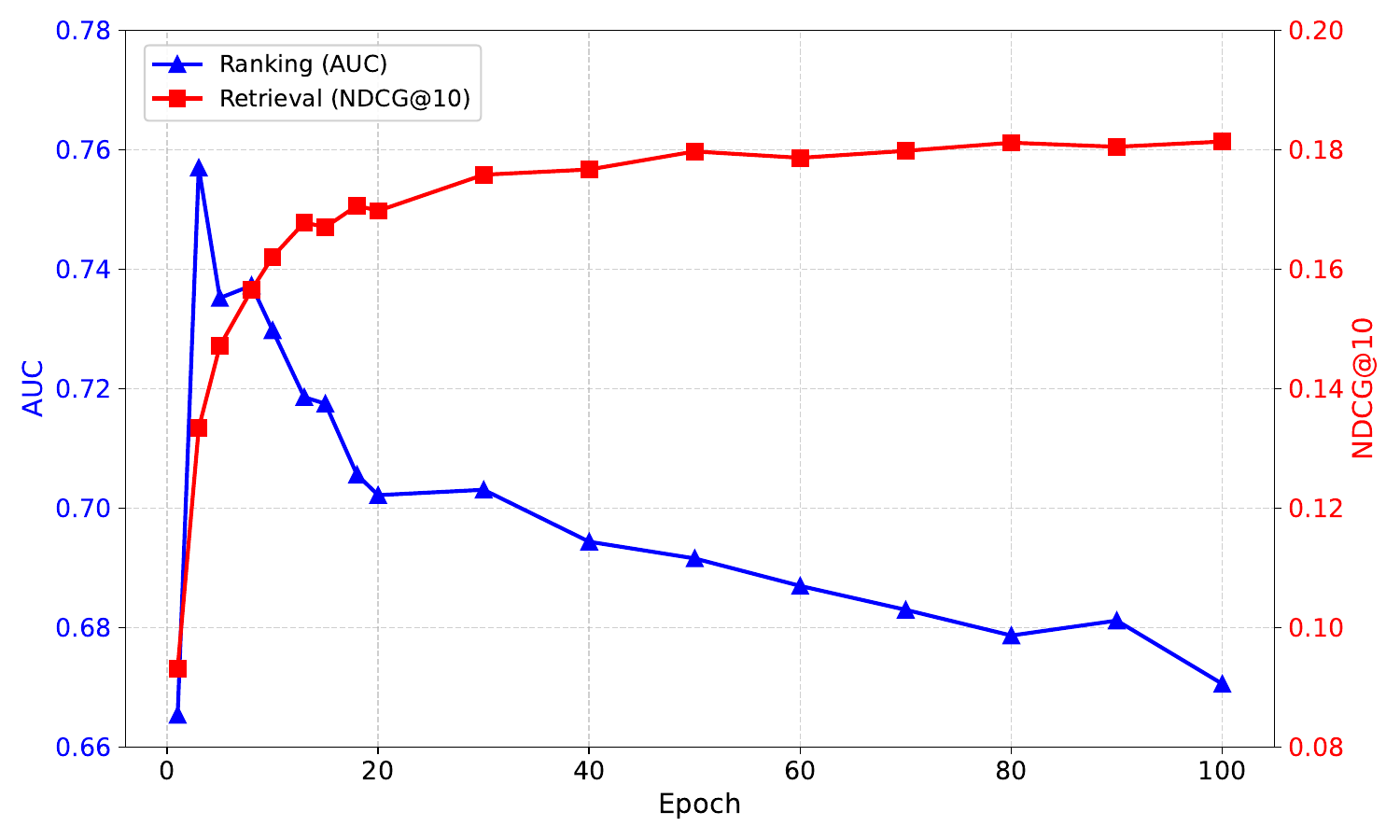}
        \caption{w/o Enhancer and Weighter}
        \label{fig:epochsubfig1}
    \end{subfigure}
    \hfill
    \begin{subfigure}{0.236\textwidth}
        \centering
        \includegraphics[width=\linewidth]{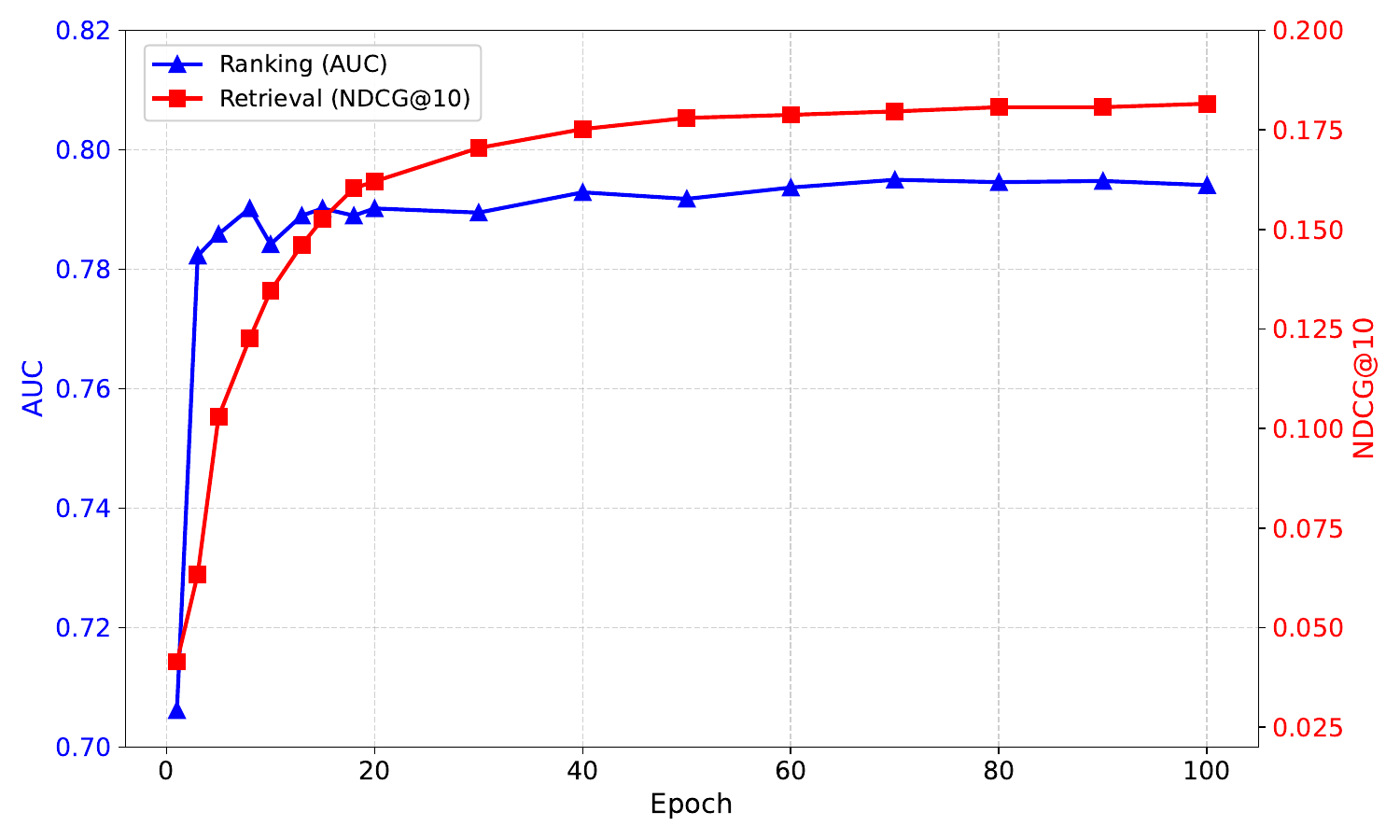}
        \caption{UniGRF-HSTU}
        \label{fig:epochsubfig2}
    \end{subfigure}
    \caption{Effect of synchronized stage optimization on unified generative framework performance.}
    \label{fig:epoch}
\end{figure}

\begin{figure}[tbp]
    \centering
    \begin{subfigure}{0.23\textwidth}
        \centering
        \includegraphics[width=\linewidth]{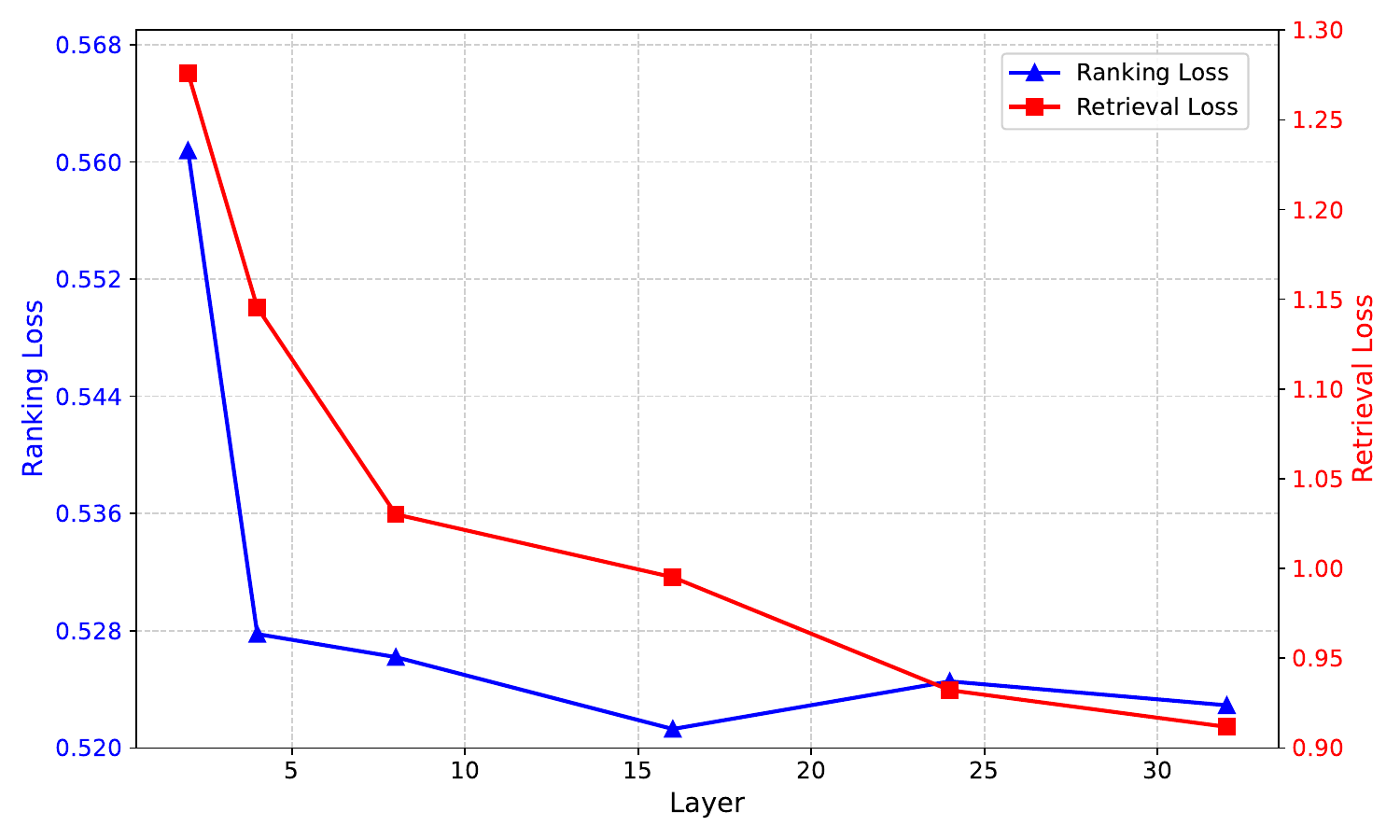}
        \caption{Loss}
        \label{fig:layersubfig1}
    \end{subfigure}
    \hfill
    \begin{subfigure}{0.23\textwidth}
        \centering
        \includegraphics[width=\linewidth]{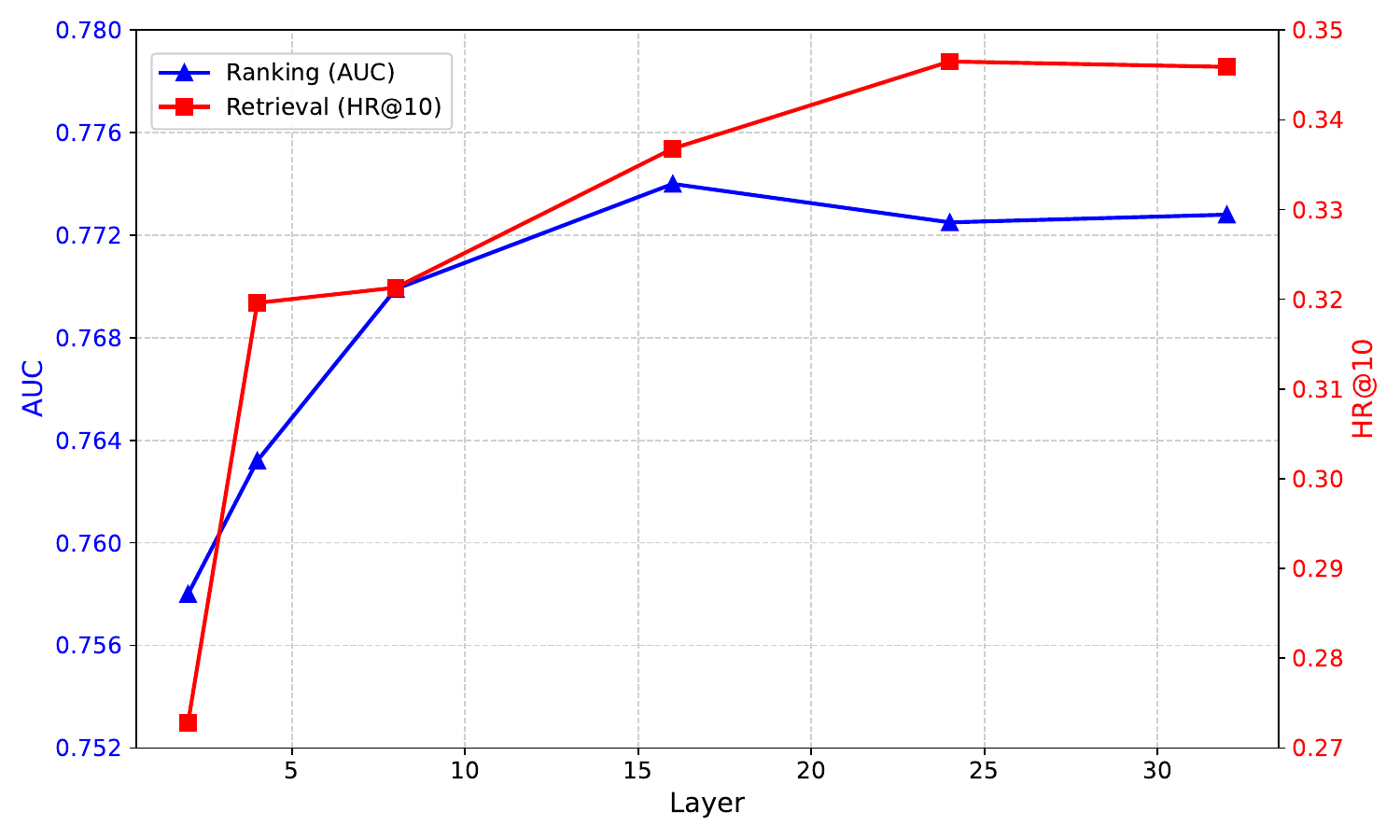}
        \caption{Performance}
        \label{fig:layersubfig2}
    \end{subfigure}
    \caption{Effect of parameter expansion on model loss and recommendation performance.}
    \label{fig:layer}
\end{figure}

\vspace{5pt}
\subsection{Further Analysis}
\subsubsection{Analysis of the Impact of Optimization on Unified Framework Performance}
To evaluate the importance of synchronized optimization between stages, we adjust the optimization speed of different stages and present the curves of recommendation performance changes over epochs in both retrieval and ranking stages, as shown in Fig.~\ref{fig:epoch}. We use the MovieLens-20M dataset for this analysis, as it exhibits the most dramatic performance changes.

Fig.~\ref{fig:epoch}(a) shows the performance change curves of the retrieval and ranking stages when using the simple addition of retrieval and ranking losses, i.e., \(\mathcal{L}^t = \mathcal{L}_{retrieval}^t + \mathcal{L}_{ranking}^t\). Unlike the continuously rising trend of the red curve in the retrieval stage, the recommendation performance in the ranking stage reaches an optimal value within several epochs and then rapidly declines.
Fig.~\ref{fig:epoch}(a) shows the performance change curves of the retrieval and ranking stages when using the simple addition of retrieval and ranking losses. Unlike the continuously rising trend of the red curve in the retrieval stage, the recommendation performance in the ranking stage (blue curve) reaches a peak after several epochs and then rapidly declines. This phenomenon is due to the lack of synchronization in training between the stages. The retrieval stage involves negative sampling, which is typically more challenging, while the training task for the ranking stage is relatively simpler. This causes the model to favor the retrieval stage during optimization, leaving the ranking task insufficiently trained. Particularly in the later stages of training, the model becomes overly biased towards the retrieval task, leading to a significant decline in the ranking stage's performance. This asynchronous optimization phenomenon limits the potential for sufficient information sharing and efficient inter-stage collaboration within the unified generative framework.

In contrast, Fig.~\ref{fig:epoch}(b) illustrates how our framework, UniGRF, achieves synchronous optimization of retrieval and ranking by adjusting the optimization pace of the two stages. The adaptive two-stage loss weighting results in a smoother performance improvement curve for the retrieval stage (red curve), while the performance of the ranking stage (blue curve) continues to improve. This demonstrates that the gradient-guided adaptive weighting strategy can monitor changes in training rates and loss convergence speeds in real time, dynamically adjusting the loss weight of each stage to ensure appropriate attention throughout the training process. By synchronously optimizing both stages, UniGRF effectively leverages the potential of the unified generative framework, significantly enhancing recommendation performance.

\subsubsection{Analysis on Scaling Law}

To investigate the performance changes of our model under parameter expansion, we adjust the number of Transformer layers within the generative architecture and observe the impact on model performance. We instantiate the Transformer block as HSTU, setting the number of layers to $\{2, 4, 8, 16, 24, 32\}$. The experimental results on the MovieLens-20M dataset for UniGRF-HSTU are presented in Fig.~\ref{fig:layer}. 

Fig.~\ref{fig:layer}(a) illustrates the change in the loss value as the model parameter size increases. It is evident that, for both retrieval and ranking tasks, as more Transformer layers are added, the loss value continues to decrease. This trend indicates that our model follows the scaling law, meaning the model can be improved by increasing the parameter size. 

In recommendation scenarios, a decrease in loss value does not always correspond to improved model performance. Therefore, we also examined how model performance changes with varying parameter sizes. As shown in Fig.~\ref{fig:layer}(b), the actual performance of the model also improves with the number of Transformer layers. This not only verifies that the decrease in loss values is consistent with the improvement in model performance, but also demonstrates that our method can more effectively capture complex patterns in the data under more complex model architectures. These results indicate that by expanding the model parameters, our model can not only reduce the loss but also significantly enhance the overall performance of retrieval and ranking, demonstrating strong application potential. This paper temporarily focuses on constructing a unified generative recommendation framework to achieve efficient collaboration and synchronized optimization between stages, and has so far only demonstrated the scalability of UniGRF on large public datasets MovieLens-20M. In future work, we will study scaling law on industrial datasets.

\section{Conclusion}
In conclusion, this paper introduced the Unified Generative Recommendation Framework (UniGRF), which addressed the inherent information loss limitation of a multi-stage paradigm by unifying separate retrieval and ranking stages of recommendation systems into a single generative model. 
UniGRF's innovative components, including the ranking-driven enhancer module and the gradient-guided adaptive weighter, enabled efficient collaboration and synchronized optimization between stages, leading to significant improvements in recommendation performance.
Through extensive experimentation on various benchmark datasets, UniGRF demonstrated its superior capability to outperform state-of-the-art baselines, validating the effectiveness of a unified framework in generative recommendation systems and offering a scalable solution that amplifies the strengths of generative models.
In the future, we plan to explore unifying more stages such as prerank and rerank within this framework.
Additionally, we will also attempt to apply this framework to specific industrial scenarios and try to achieve performance improvements by extending parameters.


\bibliographystyle{ACM-Reference-Format}
\bibliography{Sections/References}

\end{document}